\documentclass[epj]{svjour}
\usepackage{amsmath,amssymb,epsfig,bm,mathrsfs,feynmp,slashed,color,graphicx,mathtools,isotope}
\usepackage[numbers,sort&compress]{natbib}
\usepackage{filecontents}
\usepackage{tikz}
\usepackage[utf8]{inputenc}
\newcommand{\be}{\begin{equation}}
\newcommand{\ee}{\end{equation}}
\newcommand{\bea}{\begin{eqnarray}}
\newcommand{\eea}{\end{eqnarray}}
                                                               
\newcommand{\aap}{Astron.\& Astrophys.}
\newcommand{\ep}{\epsilon}

\newcommand{\D}{\Delta}
\newcommand{\vecp}{\bm p}
\definecolor{red}{rgb}{0.8,0,0}
\definecolor{orange}{rgb}{0.8,0.2,0.0}
\definecolor{blue}{rgb}{0.3,0.0,0.8}
\definecolor{violet}{rgb}{0.4,0,0.4}
\definecolor{green}{rgb}{0,0.5,0.0}

\def\apj{ApJ~}%
\def\apjl{ApJ Lett.~}%
\def\apjs{ApJS}%
\def\aap{A\&A~ }%
\def\pasa{PASA}%
\def\prc{Phys.~Rev.~C~}%
\def\prd{Phys.~Rev.~D~}%
\def\prl{Phys.~Rev.~Lett.}%
\def\nphysa{Nucl.~Phys.~A}%
\usepackage[colorlinks=true]{hyperref}
\definecolor{reddish}{rgb}{0.7,0.2,0.0}
\definecolor{blueish}{rgb}{0.1,0.1,1}
\hypersetup{
  linkcolor=reddish,          
  citecolor=blueish,           
urlcolor=magenta}            
\journalname{Eur. Phys. J. A}
\begin{document}

\title{Light clusters in dilute heavy-baryon admixed nuclear matter  }

\author{ Armen Sedrakian\inst{1,2}
}                     
\institute{
  Frankfurt Institute for Advanced Studies, Ruth-Moufang 
  str.\,\,1, D-60438 Frankfurt am Main, Germany
  \and
   Institute of Theoretical Physics, University of Wroc\l{}aw, pl. M. Borna 9, 
  50-204 Wroc\l{}aw, Poland
}
\date{Received: 1 September 2020 / Accepted: 26 September 2020\\
\textcopyright \, 
Author(s) 2020
\\
}
%
\abstract{ We study the composition of nuclear matter at
  sub-saturation densities, non-zero temperatures, and isospin
  asymmetry, under the conditions characteristic of binary neutron
  star mergers, stellar collapse, and low-energy heavy-ion collisions.
  The composition includes light clusters with mass number $A\le 4$, a
  heavy nucleus ($\isotope[56]{Fe}$), the $\Delta$-resonances, the
  isotriplet of pions, as well as the $\Lambda$ hyperon. The nucleonic
  mean-fields are computed from a zero-range density functional,
  whereas the pion-nucleon interactions are treated to leading order
  in chiral perturbation theory. We show that with increasing
  temperature and/or density the composition of matter shifts from
  light-cluster to heavy baryon dominated one, the transition taking
  place nearly independent of the magnitude of the isospin. Our
  findings highlight the importance of simultaneous treatment of light
  clusters and heavy baryons in the astrophysical and heavy-ion
  physics contexts.
} 
\maketitle
%

\section{Introduction}
\label{sec:introduction}

The formation of light clusters in dilute, warm nuclear matter is of
interest in astrophysics of binary neutron star mergers, stellar
collapse, as well as in heavy-ion physics. The details of the matter
composition are important for the accurate determination of transport
coefficients appearing in dissipative relativistic fluid dynamics as
well as the neutrino Boltzmann transport in various astrophysical
scenarios. The clustering phenomenon is also of great interest in
nuclear structure calculations (e.g. alpha-clustering) and heavy ion
collisions in laboratory experiments.


A great deal of effort during the last decade was focused on the
accurate determination of the composition of dilute nuclear matter at
finite temperatures and isospin asymmetry within a range of methods
based on the ideas of nuclear statistical
equilibrium~\cite{Sumiyoshi2008,Ducoin2008,Arcones2008PhRvC,Heckel2009,Souza2009,Typel2010,Botvina2010NuPhA,Raduta2010,Hempel2010NuPhA,Roepke2011NuPhA,Hempel2011PhRvC,Qin2012PhRvL,Hempel2012ApJ,Gulminelli2012PhRvC,Ferreira2012PhRvC,Raduta2014EPJA,Buyuk2014ApJ,Aymard2014PhRvC,Gulminelli2015PhRvC,Hempel2015PhRvC,Burrello2015PhRvC,Ropke2015PhRvC,Furusawa2016,Avancini2017,Wu2017JLTP,Zhang2017,Menezes2017PhRvC,Fortin2018PASA,Pais2018,Zhang2019PhRvC,Grams2018PhRvC,Raduta2019,Pais2019,Ropke2020,Pais2020,Mallik2020}
and virial expansion for quantum
gases~\cite{Schmidt1990,Sedrakian1998AnPhy,Horowitz2006,Mallik2008}.
The appearance of clusters leads to a range of interesting phenomena,
in particular $\alpha$-con\-den\-sation at low
temperatures~\cite{Sedrakian2006NuPhA,Wu2017JLTP,Zhang2017,Zhang2019PhRvC,Satarov2017JPhG,Satarov2019PhRvC,Furusawa2020}.
In astrophysics, light clusters and their weak interactions with
neutrinos were studied in detail in the context of stellar collapse
and supernova
physics~\cite{Fischer2014EPJA,Fischer2016EPJWC,Nagakura2019ApJS,Fischer2020}. The
electroweak interactions of leptons with baryonic matter are also of
interest in describing the transport in binary neutron star mergers,
in particular the bulk
viscosity~\cite{Alford2019PhRvC,Alford2019,ArusParticles3020034,Alford2020}
and electrical
conductivity~\cite{Harutyunyan2016PhRvC,Harutyunyan2018EPJA}.

The formation of the heavy baryons in dense and cold nuclear matter,
in particular hyperonic members of the $J^{1/2+}$ baryonic octet in
combinations with the non-strange members of baryon $J^{3/2+}$
decouplet ($\D$-resonances) has attracted attention in recent
years~\cite{Drago_PRC_2014,Cai_PRC_2015,Sahoo_PRC_2018,Kolomeitsev_NPA_2017,Li_PLB_2018,Li2019ApJ,Li2020PhRvD,Ribes_2019,Raduta2020,Sedrakian2020}.
The relativistic density functionals were successfully tuned to remove
the tension between the softening of the equation of state of dense
matter associated with the onset of the baryons and the astrophysical
observations of the massive neutron stars with masses
$2M_{\odot}$~\cite{Kolomeitsev_NPA_2017,Li_PLB_2018,Li2019ApJ}.

The motivation of this work is to explore the interplay between the
clustering and heavy-baryon degrees of freedom in dilute,
finite-temperature nuclear matter. For this purpose we set-up a model
which includes both light nuclear clusters with mass number $ A\le 4$,
a representative heavy nucleus ($\isotope[56]{Fe}$) as well as the
$\Lambda$-hyperon, the quartet of $\D$-resonances, and the isotriplet
of pions $\pi^{\pm,0}$.  Previously, hyperons were included in the
finite temperature composition of matter in stellar collapse 
and proto-neutron star
studies~\cite{Nakazato2012,Peres2013,Raduta2020}.  Pions and pion
condensation has been studied recently in the stellar context in
Refs.~\cite{Ishizuka2008,Peres2013,Colucci2014PhLB,Fore2020}. While
the light nuclear clusters have been accounted for in the low-density
envelops used in some models, a combined study of the clustering,
heavy baryons and pions is missing so far.

In this work, we extend the approach of Ref.~\cite{Wu2017JLTP} to
include heavy baryons and pions in the composition and the equation of
state of isospin asymmetrical nuclear matter. In addition to the
mean-field effects included in the previous study, we will treat also
the Pauli-blocking effects on the binding energies of the light
clusters in an approximate manner. We will focus on temperatures
$T\ge 10$~MeV, which is above the critical temperature of
Bose-Einstein condensation of $\alpha$ particles in the clustered
environment, see for further details
~\cite{Wu2017JLTP,Zhang2017,Zhang2019PhRvC,Satarov2017JPhG,Satarov2019PhRvC,Furusawa2020}. Indeed,
low temperatures disfavor the heavy baryons in low-density nuclear
matter and the problem of $\alpha$ condensation is unaffected by their
nucleation. While we include in our composition a heavy nucleus, its
effect will turn out to be minor in the parameter range studied in
this work.

The paper is organized as follows.  Section~\ref{sec:QP} extends the
formalism of the quasiparticle gas model~\cite{Wu2017JLTP} to include
heavy baryons and pions. In Sec.~\ref{sec:results} we present the
numerical results for the composition and equation of state of
matter. Section~\ref{sec:summary} provides a summary and an outlook.

\section{Formalism}
\label{sec:QP}

\subsection{Thermodynamics}

We consider matter composed of unbound nucleons, heavy baryons, light
nuclei ($A\le 4)$, $\isotope[56]{Fe}$ and pions at temperature $T$ and
baryon number density $n_B$. We assume that the charge fraction is
fixed to a value $Y_Q = n_L/n_B$, where
$n_L = (n_e-n_{e^+}) + (n_{\mu}-n_{\mu^+})$ where $n_{e}$, $n_{e^+}$,
$n_{\mu}$ and $n_{\mu^+}$ are the number densities of electrons,
positrons, muons and anti-muons.  The thermodynamical potential of the
system can be expanded into a sum of contributions of constituents
\be\label{eq:thermopotential}
\Omega (\mu_n,\mu_p, T) =\sum_{j}\Omega_{j} (\mu_{j},T),
\ee
where $j$ runs over the all elements of the composition of matter,
specifically, $j=A,Z$ for nuclei with mass number $A$ and charge $Z$,
$j=n,p$ for neutrons and protons,
$j=\{\Delta^0, \Delta^+,\Delta^{++},\Delta^{-}\}$ for $\D$-resonances,
$j=\Lambda$ for the $\Lambda$-hyperon, and $\pi^0,\pi^{\pm}$ for the
isotriplet of pions.  Here the chemical potentials of the species
$\mu_{j}$ are functions of the chemical potentials of neutrons and
protons $\mu_n$ and $\mu_p$ in ``chemical'' equilibrium with respect
to weak and strong interactions.

If a nucleus is characterized by mass number $A$ and charge $Z$ its
chemical potential is expressed as
\be\label{eq:cheq_nuclei} \mu_{A,Z} = (A-Z)\mu_n + Z \mu_p.
\ee
For the chemical potentials of heavy baryons the following relations
hold
\begin{eqnarray}
\label{eq:cheq_HB1}
  \mu_{\Lambda}&=&\mu_{\Delta^0}=\mu_n=\mu_B,\\
  \label{eq:cheq_HB2}
  \mu_{\Delta^-}&=& 2\mu_n- \mu_p=\mu_B-\mu_Q,\\
  \label{eq:cheq_HB3}
  \mu_{\Delta^+}&=& \mu_p =\mu_B+\mu_Q,\\
  \label{eq:cheq_HB4}
  \mu_{\Delta^{++}}&=& 2\mu_p- \mu_n=\mu_B+2\mu_Q,
\end{eqnarray}
where we introduced the baryon number chemical potential $\mu_B$
and the charge chemical potential $\mu_Q=\mu_p-\mu_n$. The chemical
potentials of the pions obey the following relations
\bea \label{eq:cheq_pi0}
\mu_{\pi^0}  &=& 0 ,\\
\label{eq:cheq_pi1}
\mu_{\pi^+} &=& \mu_p-\mu_n,\\
\label{eq:cheq_pi2}
\mu_{\pi^-}  &=&\mu_n-\mu_p.
\eea
The baryon number density and the charge neutrality
conditions are given by the relations
\bea
\label{eq:n_B}
n_B&=&n_n+n_p+
\sum_{c}A_cn_c \nonumber\\
&+& n_{\Delta^{++}}+n_{\Delta^{+}}  
+n_{\Delta^-}+n_{\Delta^0}+n_{\Lambda}, \\
\label{eq:Y_Q}
n_BY_Q  &=& n_p + \sum_{c}Z_cn_c\nonumber\\
&+&2n_{\Delta^{++}}+n_{\Delta^{+}}
-n_{\Delta^-}+n_{\pi^{+}}-n_{\pi^-}, 
\eea
where the $c$-summation goes over the densities of deuteron ($d$),
triton ($t$), $\isotope[3]{He}$ ($h$), $\alpha$-particle and
$\isotope[56]{Fe}$ nucleus. The latter nucleus is considered below in
its ground state, i.e., the states that are excited at finite
temperatures are neglected. The inclusion of these states will act to
enhance the fraction of this particular nucleus or other heavier
nuclei in matter, should they be included in the composition.
Equations \eqref{eq:n_B} and \eqref{eq:Y_Q} determine the two unknown
chemical potentials $\mu_{n}$ and $\mu_{p}$ at any temperature $T$ for
fixed values of $n_B$ and $Y_Q$.

The thermodynamical potential for each species can be expressed
through the densities 
\bea \label{eq:Oml}
\Omega_{j} (\mu_j, T)= - V\int_{-\infty}^{\mu_{j}} d\mu'_{j}\,\,
n_{j}(\mu'_{j}, T), \eea
where $n_{j}(\mu'_{j}, T)$ is the number density of species $j$, $V$
is the volume.

In the stellar context, the matter is charge neutral, the positive
charge of baryons being neutralized by leptons (electrons and muons).
The lepton thermodynamic potential is given by
\bea\label{Omega_l}
\Omega_{L} = - \sum_{l=e,\mu} g_lT\int\frac{d^3k}{(2\pi)^3} {\rm
  ln}\left[f_l^{-1}\left(-E_{l}(k)+\mu_l\right)\right],
\eea
where the index $l$ sumes of electrons $e$ and muons $\mu$
($\tau$-leptons can be neglected), $g_l=2$ is the degeneracy factor,
the lepton energy is given by $E_{l} = \sqrt{k^2+m_l^2}$, where $m_l$
is the lepton mass and $\mu_l$ their chemical potential and $f_l$
stands for the lepton Fermi distribution function. The lepton
density is obtained then as $n_l = \partial\Omega_L/\partial
\mu_l$. At finite temperatures a small fraction of positrons may
appear: their thermodynamical potential is obtained from
Eq.~\eqref{Omega_l} by interchanging the sign of the electron chemical
potential. To obtain the full thermodynamical potential of matter in
astrophysical contexts one needs to take into account, in addition,
the thermodynamical potential of neutrinos and anti-neutrinos. For any
fixed flavor it has the same form as Eq.~\eqref{Omega_l}, the only
difference being the degeneracy factor $g_{\nu} =1$ (as implied by the
Standard Model) and vanishingly small neutrino mass.

Having computed partial contributions $\Omega_{j}$, the thermodynamic
quantities can be obtained from the thermodynamic potential
Eq.~(\ref{eq:thermopotential}) for nuclear systems and from the
sum of Eq.~(\ref{eq:thermopotential}) and Eq.~\eqref{eq:Oml} in the
charge neutral  stellar systems. In particular, we recall that the
pressure and the entropy are given by
\begin{equation}
P= -\frac{\Omega}{V} , \quad \quad  
S = -\frac{\partial \Omega}{\partial T}.
\end{equation}

\subsection{Computing densities}

We now turn to the computation of the partial densities of
constituents. This can be done in a unified manner for quasiparticles,
resonances, and clusters using the real-time finite temperature
Green's function (hereafter GF) formalism. The density of species $j$
are directly related to the following GFs
\bea
\label{eq:Gl}
iG^<_{j}(x_1,x_2) &=& \mp \langle \psi^{\dagger}_{j}(x_2)\psi_{j}(x_1) 
\rangle,\\
\label{eq:Gg}
iG^>_{j}(x_1,x_2) &=& \langle 
\psi_{j}(x_1)\psi^{\dagger}_{j}(x_2) \rangle,
\eea
where $\psi^{\dagger}_{j}(x_1)$ and $\psi_{j}(x_1)$ are the creation
and annihilation operators of a species $j$ at the space-time point
$x_1$, the upper sign here and below refers to fermions, the lower --
to bosons.  The time-arguments of the GF are located on different
branches of the Schwinger-Keldysh time-contour with $t_2< t_1$ in
\eqref{eq:Gl} and $t_2> t_1$ in \eqref{eq:Gg}. The Fourier transforms
of GFs in \eqref{eq:Gl} and \eqref{eq:Gg} with respect to the argument
$x_1-x_2$ are related to the occupation numbers and the spectral
function $ S_j(\omega,\vecp)$ as
\bea \label{eq:GFourier1}
-i G^<_{j}(\omega, \vecp)  &=& \pm S_j(\omega,\vecp) f(\omega),\\
\label{eq:GFourier2}
i G^>_{j}(\omega, \vecp) &=& S_j(\omega,\vecp) [1\mp f(\omega)],
\eea
where $f_{j}(\omega)$ is either Bose or Fermi distribution function
depending on the spin of the $j$-species.  From
Eqs. \eqref{eq:GFourier1} and \eqref{eq:GFourier2} in follows that 
\bea
i G^>_{j}(\omega, \vecp) -i G^<_{j}(\omega, \vecp) = S_j(\omega,\vecp).
\eea
At this point it is convenient to establish the connection to the
advanced (A) and retarded (R) GFs
\bea
 [G^{R/A}_{j}(\omega, \vecp)]^{-1} =
\omega -\ep_{\vecp} - \Sigma^{R/A}(\omega, \vecp),
\eea
where $\ep_{\vecp}$ the energy of particle in the non-interacting
theory and $\Sigma^{R/A}(\omega, \vecp)$ are the retarded/advanced
self-energies that are commonly evaluated in the equilibrium
theory. If we use the identity
\bea
G^>_j(\omega, \vecp) - G^<_j(\omega, \vecp) =
G^{R}_j(\omega, \vecp) - G^{A}_j(\omega, \vecp),
\eea
the spectral function takes the form 
\be \label{eq:SFunction}
S_{j}(\omega,\vec p) = 
\frac{\Gamma_{j}(\omega,\vec p)}
{[\omega-E_{j}(\omega,\vec p)]^2
+\Gamma_{j}^2(\omega,\vec p)/4},
\ee
where $E_{j}(\omega,\vecp)$ is the quasiparticle energy
and $\Gamma_{j}(\omega,\vecp) 
= -2 {\rm Im}\Sigma_{j}(\omega, \vecp)$ is the spectral 
width. The  quasiparticle energy is given by
\bea \label{eq:QP_E}
E_{j}(\omega,\vecp) &=& \frac{p^2}{2m_j}+E^0_j +
{{\rm Re}\,\Sigma_{j}(\omega,\vecp)}-\mu_{j},
\eea
where $m_j$ is the mass, $E^0_j$ is the vacuum binding energy of 
the nucleus $j=(A,Z)$, which vanishes for baryonic
quasiparticles. From the definition \eqref{eq:Gl} it follows
that
\bea\label{eq:density} n_{j} &=&- i g_{j}\int\frac{d\omega d\vecp}{(2\pi)^4}
G^<_{j}(\omega,\vecp), \eea
where $g_{j}$ is the degeneracy factor.

Thus, we have obtained a closed set of equations which consists of 
Eqs.~\eqref{eq:n_B} and \eqref{eq:Y_Q} for the two unknowns $\mu_n$
and $\mu_p$ at fixed $n_B$, $Y_Q$ and $T$, whereby the densities of
constituents are computed from Eqs.~\eqref{eq:GFourier1},
\eqref{eq:SFunction} \eqref{eq:QP_E} and \eqref{eq:density}.  These
equations still contain unspecified self-energies of the constituents,
which depend on the modeling of the interactions in the system.  We
turn now to this problem.

\subsection{Self-energies}

We assume that in the dilute limit of interest the unbound baryons are
well-defined quasiparticles and the imaginary part of their
self-energy vanishes; this implies that their spectral function is a
delta-function 
\bea\label{eq:S_nucleons}
S_{j}(\omega,\vecp)  = 2\pi\delta(\omega -\ep_{\vecp,j}
- {\rm Re}\Sigma_j^{R/A}(\omega, \vecp)),
\eea
where
\bea
\ep_{\vecp,j} = \frac{p^2}{2m_j^*} -\mu_j
\eea
with $m_j^*$ and $\mu_j$ being the effective mass and the chemical potential.
Furthermore, the nucleon self-energy is approximated by the effective
masses of neutrons $m^*_n$ and protons $m^*_p$ which depend on the
baryon and charge density (or $n_B$ and $Y_Q$) but are independent of
temperature.  In the numerical work, we use the Skyrme functional
parametrization given by Eq. (15) of Ref.~\cite{Davesne2016} for that
purpose. The spectral functions of the light clusters are approximated
also by their quasiparticle limit
\be 
S_{j}(\omega,\vec p) = 2\pi\delta\left(\omega-\frac{p^2}{2M}
-E^0_j -{{\rm Re}\Sigma_{j}}+\mu_{j}
\right),
\ee
where $E^0_j$ is the vacuum binding energy of cluster $j$, ${\rm
  Re}\Sigma_{j}$ is its self-energy. The effective mass of a cluster
is constructed as $M= (A-Z)m_n^* + Zm_p^*$. Finally, for the $\Lambda$
hyperon and $\D$-resonances we use again Eq.~\eqref{eq:S_nucleons}
with their vacuum masses and neglect the narrow (118 MeV) width of the
$\D$-resonance and self-energy corrections.  With these
  approximations the energy integral in Eq.~(\ref{eq:density}) is
  trivial and one is left with the momentum phase-space integration.
  In the case of pions, we include the leading contribution to the
  pion self-energy in chiral perturbation
  theory~\cite{Oller2010,Colucci2014PhLB}, which arises from their
  coupling to neutrons and protons, specifically Eq.~(3.4) and (3.5)
  in Ref.~\cite{Oller2010}.

The binding energies of clusters are functions of density and
temperature in general. The nuclear environment influences the binding
energies through phase space occupation (Pauli-blocking). To take this
into account, we use the results of the solutions of in-medium two-body
Bethe-Salpeter and three-body Faddeev equations in dilute nuclear
matter given in Ref.~\cite{SedrakianClark2006}. These solutions are
fitted by the following procedure: (a) first we determine the critical
value of the inverse temperature $\beta$ for which a cluster
disappears via the formula:
$\beta_{\rm cr} [{\rm MeV}^{-1}]= 0.07835 + 0.00185~(n_0/n_B)$, where
$n_0 = 0.16$ fm$^{-3}$,
which is assumed to be universally independent of $A$ and $Y_Q$, and (b) the
in-medium binding energies $B_j(n_B, T)$ are obtained via a linear fit given by
\bea
B_j(n_B, T) = E^0_j\left[ 1-\frac{\beta}{\beta_{\rm cr}(n_0/n_B)}\right].
\eea
Then the
spectral function \eqref{eq:S_nucleons} takes the form 
\bea\label{eq:S_nucleons2}
S_{j}(\omega,\vecp)  \simeq 2\pi\delta\left(\omega-\frac{p^2}{2M} -B_j+\mu_j^*\right),
\eea
where any contribution to the self-energy beyond the modifications of
the binding energy is energy and momentum independent and, thus, can
be absorbed in the chemical potential $\mu_j^*$.

\begin{figure}[t] 
\includegraphics[width=1.\hsize]{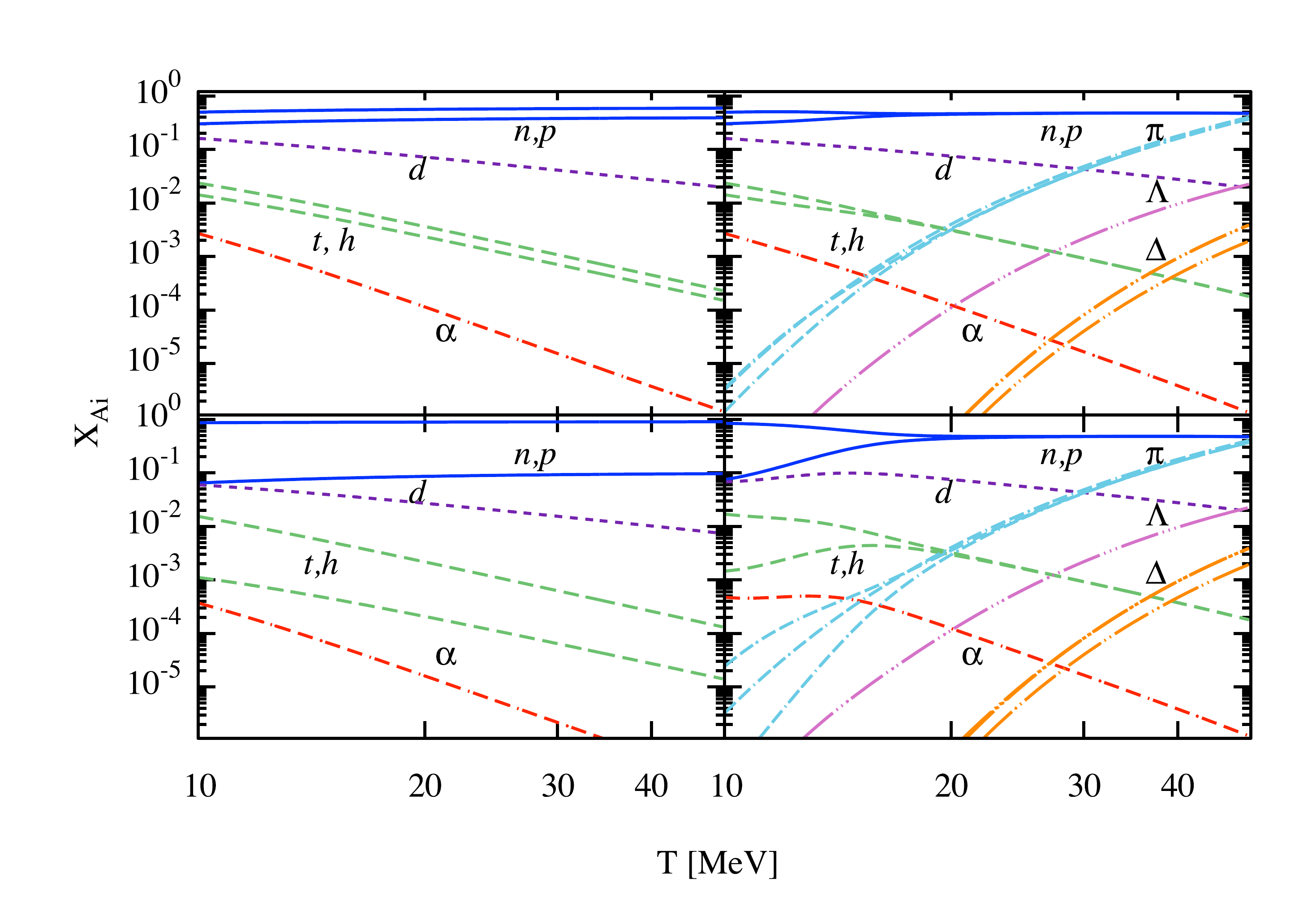}
\caption{Dependence of the mass fractions of the particles in dilute
  nuclear matter on temperature at constant density $n_B/n_0 =
  10^{-2}$. The top and lower panels correspond to charge fractions
  $Y_Q = 0.4$ and 0.1. The left and right panels correspond to the
  cases containing nucleons and light clusters only and the full
  composition, respectively. The composition includes neutrons and
  protons (solid lines), deuterons (short-dashed), triton and helium
  (long-dashed), $\alpha$-particles (dash-dotted), $\Delta$ resonances
  (dash-double-dot), $\Lambda$-hyperon (dash-triple-dot), and pions
  (double-dash-dot). The mass fraction of $\isotope[56]{Fe}$ is not
  visible on the figure's scale.}
\label{fig:X-T}
\end{figure}

\section{Results}
\label{sec:results}

The system of Eqs.~\eqref{eq:n_B} and \eqref{eq:Y_Q} was solved
simultaneously for unknown chemical potentials~$\mu_n$ and $\mu_p$ at
fixed temperature $T$, baryon number density $n_B$ and charge fraction
$Y_Q$. We consider two values of the latter parameter $Y_Q = 0.1$,
which is characteristic to binary neutron star mergers, and $Y_Q =
0.4$ which is characteristic to stellar collapse.

Figure~\ref{fig:X-T} shows the {\it mass fraction} $X_j = A_j
n_j/n_B$, where $A_j$ is the mass number of a constituent, as a
function of temperature in cases (a) nucleons and clusters only and
(b) nucleons, clusters, heavy baryons and pions, for $Y_Q=0.1$ and 0.4
at fixed $n_B/n_0=10^{-2}$, where $n_0=0.16$ fm$^{-3}$ is the nuclear
saturation density. The mass fraction of $\isotope[56]{Fe}$ is not
visible on figure's scale.  It is seen that nucleons are the dominant
  component at all temperatures, but there is a change in the
  composition of matter with respect to the remaining constituents
  with increasing temperature.  For temperatures $T \ge 30$ MeV the
  dominant mass fraction is in the heavy baryons, whereas at lower
  temperatures the clusters are the dominant component. Note also that
  the inclusion of heavy baryons and pions reduces the isospin
  asymmetry in the neutron and proton components and, as a
  consequence, the helion and triton abundances are much closer to
  each other in this case. A previous study of hyperon
    abundances at finite temperatures in Ref.~\cite{Fortin2018PASA}
    finds that the hyperon fraction exceeds $10^{-4}$ at density
    $n_B/n_0=10^{-2}$ for temperatures $ T\ge 40$~MeV. According to
    Fig.~\ref{fig:X-T} this occurs in our model for $ T\ge
    20$~MeV. This difference may be a consequence of different
    treatment of nuclear interactions and different compositions
    allowed in the models.  Ref.~\cite{Menezes2017PhRvC} finds that
    $\Lambda$ hyperon fraction stays below $10^{-7}$ for temperatures
    up to $14$ MeV in the inhomogeneous ``pasta'' phases of supernova
    matter independent of the value of $Y_Q$, which is consistent with
    present results.  

\begin{figure}[tbh] 
 \includegraphics[width=1.\hsize]{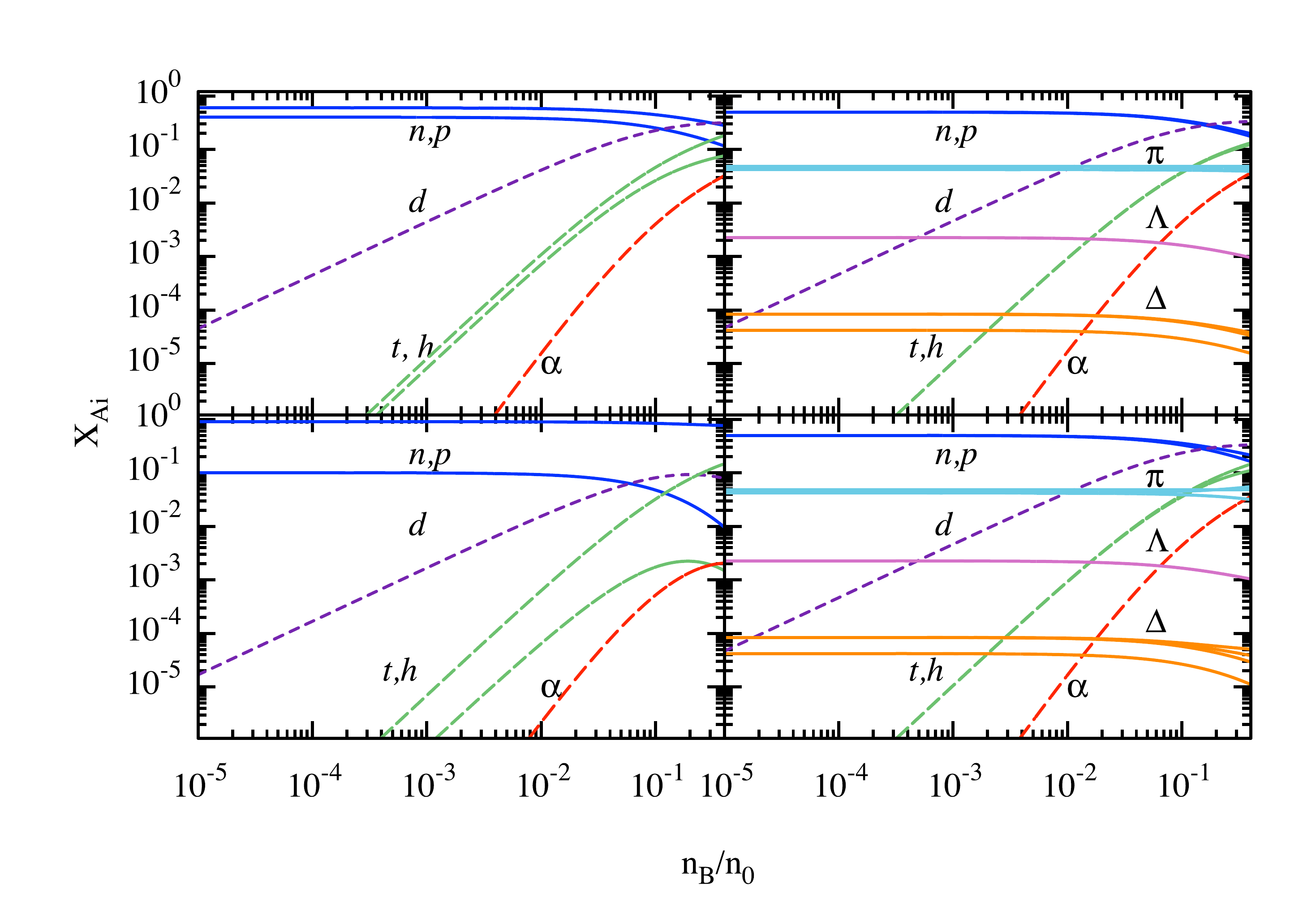}
 \caption{ Dependence of the mass fractions of the particles in dilute nuclear matter on density for $T=30$~MeV. The top and lower panels correspond to charge fraction $Y_Q = 0.4$ and 0.1 and the left and right panels correspond to the cases containing only nucleons and light clusters and the full composition,
   respectively. The composition includes neutrons and protons (solid
   lines), deuterons (short-dashed), triton and helium (long-dashed),
   $\alpha$-particles (dash-dotted), $\Delta$ resonances
   (dash-double-dot), $\Lambda$-hyperon (dash-triple-dot), and pions
   (double-dash-dot). In the right figure, the clusters disappear for
   $n_B/n_0 \ge 9\times 10^{-2}$ (shaded area) due to the
   Pauli-blocking of the phase-space. The mass fraction of
   $\isotope[56]{Fe}$ is not visible on the figure's scale.}
\label{fig:X-rho1}
\end{figure}
\begin{figure}[tbh] 
  \includegraphics[width=1.\hsize]{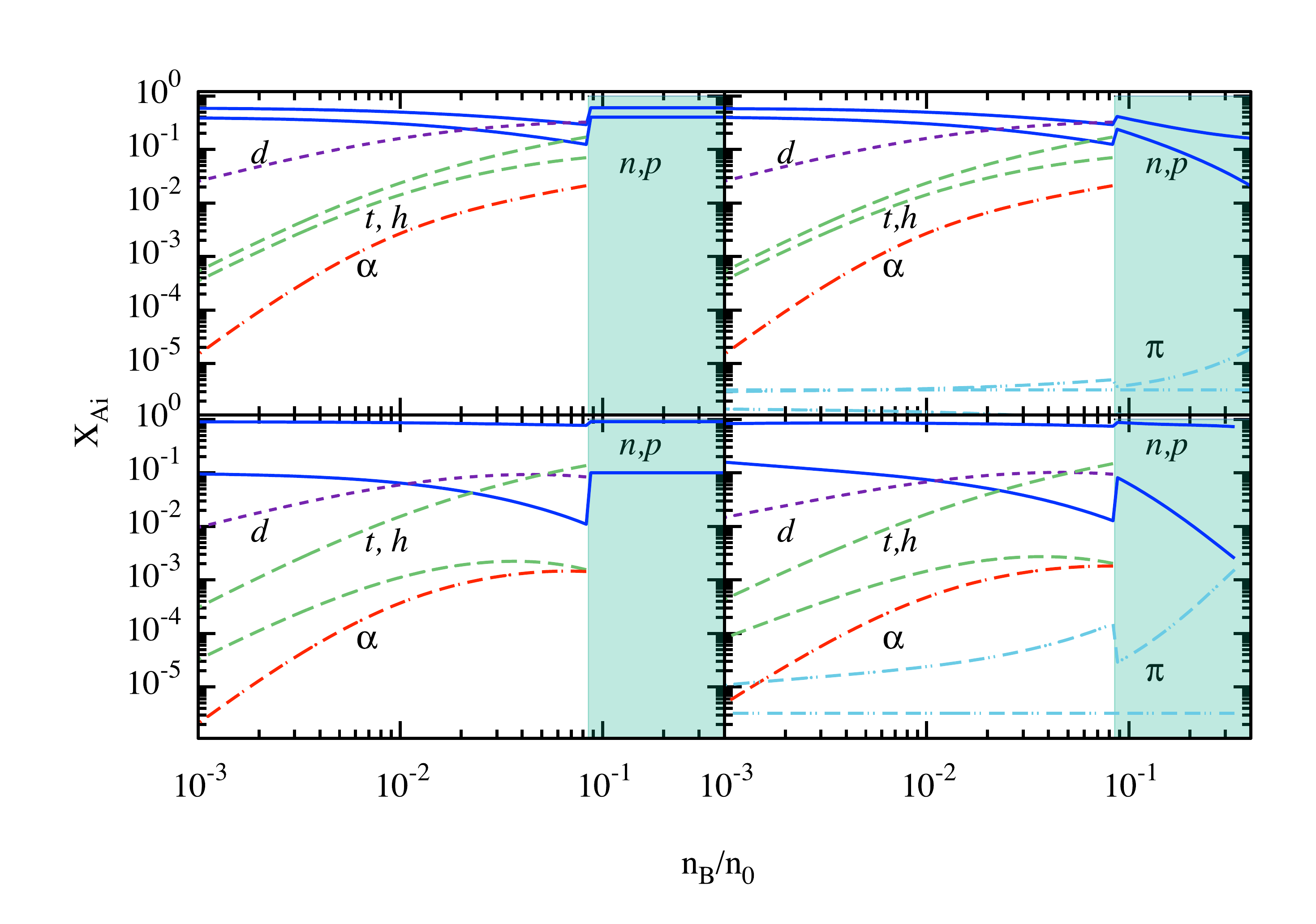}
   \caption{ Same as in Fig.~\ref{fig:X-rho1} but for  $T=10$~MeV.}
  \label{fig:X-rho2}
  \end{figure}
  Figures \ref{fig:X-rho1} and \ref{fig:X-rho2} show the mass
  fractions mass fraction $X_j $ at two fixed temperatures $T=30$~MeV
  and $T=10$~MeV and varying density. It is seen that the abundances
  of the nucleons, heavy baryons, and pions are insensitive to the
  density, whereas the cluster abundances increase as the density
  increases.  In other words, the increase in the nucleonic density at
  a fixed temperature is accommodated by the system by increasing the
  number of the light clusters, whereas the fractions of neutrons and
  protons remain constant in a wide density range. Since the heavy
  baryon fraction are determined by their “chemical” equilibrium with
  respect to neutrons and protons via the relations
  \eqref{eq:cheq_HB1}-\eqref{eq:cheq_HB4}, their fractions stay
  constant with the density as well. The same applies also to pion
  fractions, which are likewise related to proton and neutron
  concentrations via Eqs.~\eqref{eq:cheq_pi1} and \eqref{eq:cheq_pi2}.
  The reduction of isospin asymmetry among neutrons and protons
  mentioned above is seen here as well. Note that the Pauli-blocking
  at $T=30$~MeV is ineffective within the density range considered,
  but its effect is seen in the right panels of Fig.~\ref{fig:X-rho2}
  corresponding to $T=10$~MeV. It is seen that $n_B/n_0 \simeq 0.1$
  the clusters abruptly disappear as a consequence of $B_j(n_B, T)\to
  0$ and there appears a jump in the density of nucleons. Note that
  our Pauli blocking factor does not dependent on the momentum of the
  cluster with respect to the medium. In general, it does, so that the
  phase space vanishes with increasing the density more smoothly: the
  clusters with the lower-momenta are eliminated first, while those
  with high-momenta remain intact.  It is also seen that the pion mass
  fraction undergoes at the same point an abrupt change, clearly
  visibly for $Y_Q =0.1$. Finally, note that at this temperature, the
  heavy baryon fractions are too low to be relevant.  To assess if
  there is a phase transition (and if so, to find its order) a
  detailed study of the thermodynamic functions of matter at the point
  of the dissolution of clusters is needed.  In a similar study of
  Ref.~\cite{Typel2010}, which used non-linear fits to the binding
  energies of clusters, the transition is found to be less abrupt.

\begin{figure}[t] 
\includegraphics[width=1.\hsize]{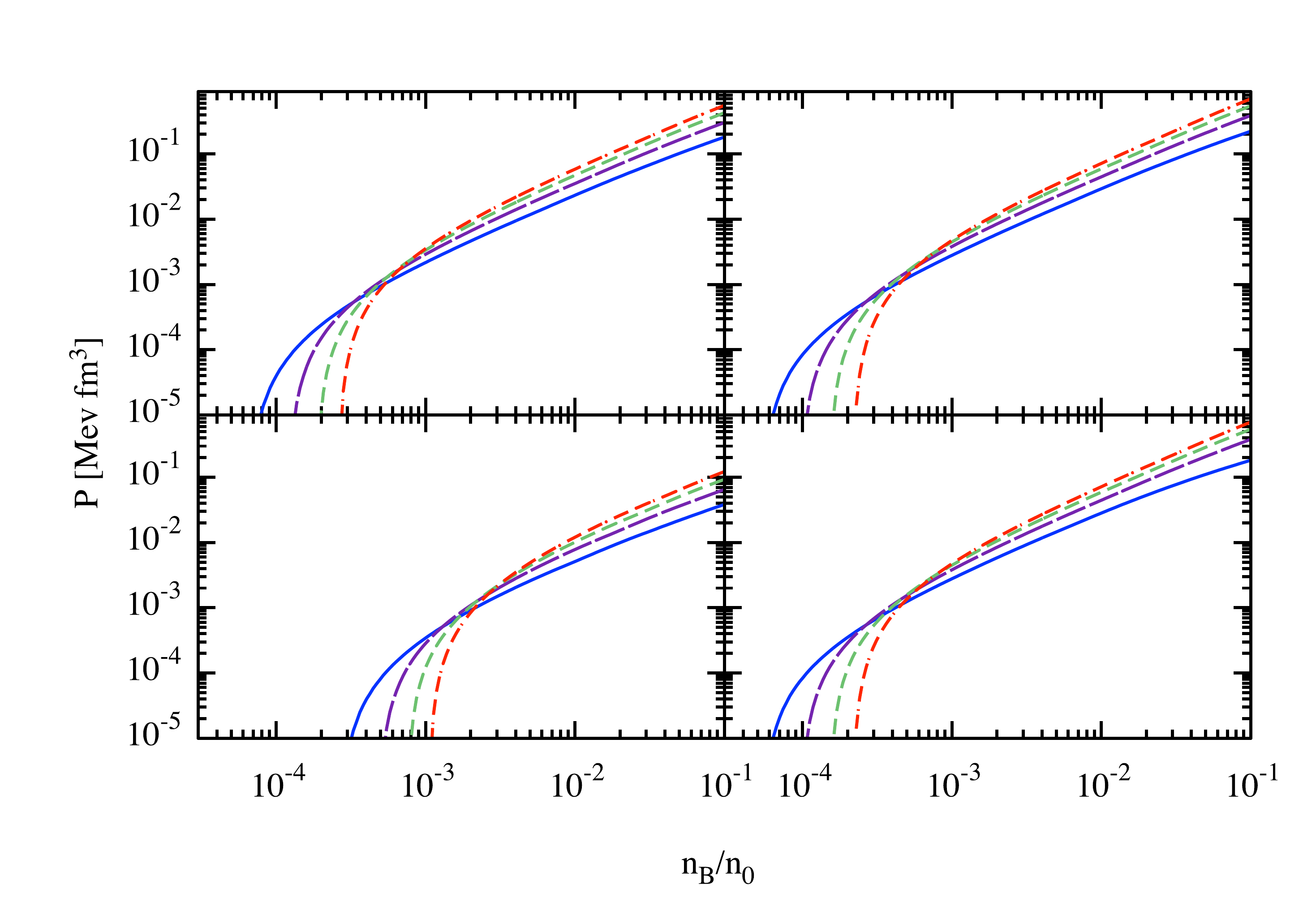}
\caption{ Pressure as a function of normalized density $n_B/n_0$
  for temperature values (in MeV) $T=20$ (solid lines), 30 (long-dashed), 40
  (short-dashed), and 50 (dash-doted). The upper panels correspond to
  $Y_Q = 0.4$ and the lower ones - to $Y_Q = 0.1$. The composition in
  panels is as in Fig.~\ref{fig:X-T}.}
\label{fig:EOS}
\end{figure}

Figure~\ref{fig:EOS} shows the pressure as a function of the
normalized density for temperature values $T=20,$ 30, 40, and 50~MeV
for two values of charge fraction $Y_Q = 0.1$ and 0.4.  The main effect
caused by the onset of heavy baryons and pions is the more symmetric
appearance of the nucleonic component for $Y_Q = 0.1$, already
observed in Fig.~\ref{fig:X-rho1}, which leads to
pressure values that are similar to those for the case $Y_Q =
0.4$.

\section{Summary and outlook}
\label{sec:summary}

The composition of warm dilute nuclear matter was computed including
simultaneously light clusters with $A\le 4$, a representative heavy
nucleus ($\isotope[56]{Fe}$), heavy baryons ($\Lambda$'s and
$\Delta$'s) and pions. We find that with increasing temperature the
mass fraction shifts from light clusters to heavy baryons, whereby the
nucleons remain the dominant component within the parameter range
considered. The heavy nucleus $\isotope[56]{Fe}$ does not play a
significant role at temperatures $T\ge 10$ MeV, but is known to
suppress strongly the abundances of light clusters at low temperatures
of the order 1 MeV~\cite{Wu2017JLTP,Furusawa2020}. The addition of
heavy baryons and pions makes the nucleonic component more isospin
symmetric and, as a consequence, the cluster abundances become less
sensitive to the value of the isospin asymmetry. At low temperatures
$T\simeq 10$~MeV, the phase-space occupation strongly suppresses the
cluster abundances for densities $n_B/n_0 \ge 0.1$ due to the Pauli
blocking, as expected.

The rich composition of matter in the parameter range considered may
have interesting implications in astro\-phy\-sics of compact star
mergers, stellar collapse as well as heavy-ion collisions. The
transport studies of hadronic matter and its coupling to leptons in
these contexts need to include the additional degrees of freedom shown
to be important in the composition of matter.

\section*{Acknowledgements}
The support through the European COST Actions ``PHA\-ROS" (CA16214)
and by the Deutsche Forschungsgemeinschaft (Grant No. SE 1836/5-1) is
gratefully acknowledged.   This research has made use of NASA's
Astrophysics Data System Bibliographic Services.

\bibliographystyle{JHEP}                                                          
\providecommand{\href}[2]{#2}\begingroup\raggedright\endgroup

\end{document}